\documentstyle[12pt,amsfonts]{article}

\def\noi{\noindent}

\def\beq#1{\begin{equation}\label{#1}}
\def\eeq{\end{equation}}
\def\ber#1{\begin{eqnarray}\label{#1} \nqq}%   left alignment
\def\eer{\end{eqnarray}}

\newcommand{\bear}[1]{\begin{eqnarray}\label{#1}}
\newcommand{\ear}{\end{eqnarray}}

\renewcommand{\theequation}{\arabic{section}.\arabic{equation}}
\catcode`\@=11 \@addtoreset{equation}{section}\catcode`\@=12

\newcommand{\N}{ {\mathbb N} }
\newcommand{\R}{ {\mathbb R} }
\newcommand{\C}{ {\mathbb C} }

\newcommand{\sh}{\sinh}
\newcommand{\ch}{\cosh}

\newcommand{\sign}{\mathop{\rm sign}\nolimits}

\newcommand{\eps}{\varepsilon}
\newcommand{\tri}{\triangle}

\newcommand{\p}{\partial}
\newcommand{\nn}{\nonumber}

\newcommand{\fnm}{\footnotemark}
\newcommand{\fnt}{\footnotetext}

\begin{document}

\begin{flushright}        IGC-PFUR-03/2002  \end{flushright}

\vspace{15pt}

\begin{center}
\large\bf
%\\
COMPOSITE S-BRANE SOLUTIONS \\
RELATED TO TODA-TYPE SYSTEMS
\\[15pt]

\normalsize\bf V.D. Ivashchuk\fnm[1]\fnt[1]{ivas@rgs.phys.msu.su},

\it Center for Gravitation and Fundamental Metrology,
VNIIMS, 3/1 M. Ulyanovoy Str.,
Moscow 119313, Russia

\it Institute of Gravitation and Cosmology,
Peoples' Friendship University of Russia,
6 Miklukho-Maklaya St., Moscow 117198, Russia

\end{center}
\vspace{15pt}

\small\noi

\begin{abstract}

Composite $S$-brane solutions
in multidimensional gravity with scalar fields and
fields of forms related to Toda-like systems
are presented. These solutions are defined on a product
manifold $\R_{*} \times M_1 \times  \ldots  \times M_n$,
where $\R_{*}$ is a time manifold, $M_1$ is an Einstein manifold
and $M_i$ ($i > 1$) are Ricci-flat manifolds.
Certain examples of $S$-brane solutions  related to
$A_1 + ... + A_1$, $A_m$  Toda
chains and those with "block-orthogonal" intersections
(e.g.  $SM$-brane solutions) are singled out.
Under certain restrictions imposed
a  Kasner-like asymptotical behaviour of the
solutions  is shown.

\vspace{10cm}

\end{abstract}

\pagebreak

\normalsize

%%%%%%%%%%%%%%%%%%%%%%%%%%%%%%%%%%%%%%%%%%%%%%%%%%%%%%%%%%%%%%%%
\section{Introduction}
%%%%%%%%%%%%%%%%%%%%%%%%%%%%%%%%%%%%%%%%%%%%%%%%%%%%%%%%%%%%%%%%

Currently, there is a  certain interest  to  so-called
$S$-brane solutions \cite{S1}, i.e. space-like analogues of $D$-branes,
see also \cite{S2,S3,S4,S5} and references therein.

We remind that in  perturbative string theory
$D$-branes \cite{Polc} are  hypersurfaces where open strings end
and the Dirichlet boundary conditions along transverse spacelike directions
are imposed. Alternatively,  $D$-branes may be also described  as classical
solutions in  supergravity theories. Open strings can also obey Dirichlet
boundary conditions along time-like  directions that gives   rise to
space-like  analogues of $D$-branes, i.e. $S$-branes.
(For Euclidean brane solutions in
type $II$ string theories see also \cite{Hull1,Hull2}.)

It is well-known that $D$-branes play an important role
in studying non-perturbative aspects of string/M theory and the AdS/CFT
duality \cite{Ads1,Ads2,Ads3}. Analogously
$S$-branes are expected to play the role of $D$-branes in realizing $dS/CFT$
duality \cite{Str} in string/M theory.

Nevertheless, from pure mathematical point of view $S$-brane solutions
\cite{S1,S2,S3,S4,S5} are not new ones but were considered (mostly)
in some earlier publications on string cosmology,
see \cite{BF,LOW,LMPX,LMP,LP1} etc.
\fnm[2]\fnt[2]{The solutions from
\cite{LMPX} contain the spacelike brane
solutions subsequently obtained by
Strominger and  Gutperle \cite{S1} and some other authors.}
They are also special cases of more general
exact solutions found in certain publications
of Moscow group, see \cite{GrIM,BGIM,IMJ,GM,IK,IMtop}
and references therein.
These publications
contain a lot of exact solutions defined on product of several Ricci-flat or
Einstein spaces of arbitrary signatures. Among them
generalized $S$-brane solutions governed by harmonic function
with brane  intersections corresponding to hyperbolic algebras were
considered \cite{IMBl}. In \cite{IMb1}
a general class of cosmological solutions with composite
$S$-branes exhibiting an oscillating behaviour near the singularity
was described, e.g. using the billiard representation near
the singularity.

The aim of this paper is twofold. Firstly, we single out a family of
cosmological solutions with composite $S$-branes from
general Toda-like cosmological-type  solutions suggested in \cite{IK}.
Secondly, for special  intersections (e.g. "orthogonal" ones) we get
a subfamily of $S$-brane solutions, containing
the main part of solutions from
\cite{S1,S2,S3,S4,S5} as special cases.

The paper is organized as follows. In Section 2
cosmological-type solutions related to Toda-type systems from \cite{IK}
are presented (in a more or less condensed manner).
In Section 3 we single out  composite  $S$-brane solutions of general
(Toda-like) form. Here we also consider special subclasses
of $S$-brane solutions, e.g. with "orthogonal", "block-ortogonal"
and $A_m$ intersection rules.
In Section 4  certain examples
of solutions in $D= 11$ supergravity  (describing $SM$-branes)
are singled out.
Section 5 is devoted to asymptotical Kasner-like behaviour
of $S$-brane solutions.

%%%%%%%%%%%%%%%%%%%%%%%%%%%%%%%%%%%%%%%%%%%%%%%%%%%%%%%%%%%%%%%%

\section{Cosmological-type solutions related to Toda-type systems}
%%%%%%%%%%%%%%%%%%%%%%%%%%%%%%%%%%%%%%%%%%%%%%%%%%%%%%%%%%%%%%%%

\subsection{The model}

We consider a  model governed by the action
\beq{1.1}
S_g=\int d^Dx \sqrt{|g|}\biggl\{R[g]-h_{\alpha\beta}g^{MN}\p_M\varphi^\alpha
\p_N\varphi^\beta-\sum_{a\in\tri}\frac{\theta_a}{n_a!}
\exp[2\lambda_a(\varphi)](F^a)^2\biggr\}
\eeq
where $g=g_{MN}(x)dx^M\otimes dx^N$ is a metric,
$\varphi=(\varphi^\alpha)\in\R^l$ is a vector of scalar fields,
$(h_{\alpha\beta})$ is a  constant symmetric
non-degenerate $l\times l$ matrix $(l\in \N)$,
$\theta_a=\pm1$,
\beq{1.2a}
F^a =    dA^a
=  \frac{1}{n_a!} F^a_{M_1 \ldots M_{n_a}}
dz^{M_1} \wedge \ldots \wedge dz^{M_{n_a}}
\eeq
is a $n_a$-form ($n_a\ge1$), $\lambda_a$ is a
1-form on $\R^l$: $\lambda_a(\varphi)=\lambda_{\alpha a}\varphi^\alpha$,
$a\in\tri$, $\alpha=1,\dots,l$.
In (\ref{1.1})
we denote $|g| =   |\det (g_{MN})|$,
\beq{1.3a}
(F^a)^2_g  =
F^a_{M_1 \ldots M_{n_a}} F^a_{N_1 \ldots N_{n_a}}
g^{M_1 N_1} \ldots g^{M_{n_a} N_{n_a}},
\eeq
$a \in \tri$. Here $\tri$ is some finite set.
For pseudo-Euclidean metric of signature $(-,+, \ldots,+)$
all $\theta_a = 1$.

\subsection{Solutions with $n$ Ricci-flat spaces}

Let us consider a family of
solutions to field equations corresponding to the action
(\ref{1.1}) and depending upon one variable $u$
\cite{IK}. These solutions are defined on the manifold
\beq{1.2}
M =    (u_{-}, u_{+})  \times
M_1  \times M_2 \times  \ldots \times M_{n},
\eeq
where $(u_{-}, u_{+})$  is  an interval belonging to $\R$.
The solutions read \cite{IK}
\bear{1.3}
g= \biggl(\prod_{s \in S} [f_s(u)]^{2 d(I_s) h_s/(D-2)} \biggr)
\biggr\{ \exp(2c^0 u + 2 \bar c^0) w du \otimes du  + \\ \nn
\sum_{i = 1}^{n} \Bigl(\prod_{s\in S}
[f_s(u)]^{- 2 h_s  \delta_{i I_s} } \Bigr)
\exp(2c^i u+ 2 \bar c^i) g^i \biggr\}, \\ \label{1.4}
\exp(\varphi^\alpha) =
\left( \prod_{s\in S} f_s^{h_s \chi_s \lambda_{a_s}^\alpha} \right)
\exp(c^\alpha u + \bar c^\alpha), \\ \label{1.5}
F^a= \sum_{s \in S} \delta^a_{a_s} {\cal F}^{s},
\ear
$\alpha=1,\dots,l$; .
In  (\ref{1.3})  $w = \pm 1$,
$g^i=g
_{m_i n_i}^i(y_i) dy_i^{m_i}\otimes dy_i^{n_i}$
is a Ricci-flat  metric on $M_{i}$, $i=  1,\ldots,n$,
\beq{1.11}
\delta_{iI}=  \sum_{j\in I} \delta_{ij}
\eeq
is the indicator of $i$ belonging
to $I$: $\delta_{iI}=  1$ for $i\in I$ and $\delta_{iI}=  0$ otherwise.

The  $p$-brane  set  $S$ is by definition
\beq{1.6}
S=  S_e \sqcup S_m, \quad
S_v=  \sqcup_{a\in\tri}\{a\}\times\{v\}\times\Omega_{a,v},
\eeq
$v=  e,m$ and $\Omega_{a,e}, \Omega_{a,m} \subset \Omega$,
where $\Omega =   \Omega(n)$  is the set of all non-empty
subsets of $\{ 1, \ldots,n \}$.
Here and in what follows $\sqcup$ means the union
of non-intersecting sets. Any $p$-brane index $s \in S$ has the form
\beq{1.7}
s =   (a_s,v_s, I_s),
\eeq
where
$a_s \in \tri$ is colour index, $v_s =  e,m$ is electro-magnetic
index and the set $I_s \in \Omega_{a_s,v_s}$ describes
the location of $p$-brane worldvolume.

The sets $S_e$ and $S_m$ define electric and magnetic
$p$-branes, correspondingly. In (\ref{1.4})
\beq{1.8}
\chi_s  =   +1, -1
\eeq
for $s \in S_e, S_m$, respectively.
In (\ref{1.5})  forms
\beq{1.9}
{\cal F}^s= Q_s
\left( \prod_{s' \in S}  f_{s'}^{- A_{s s'}} \right) du \wedge\tau(I_s),
\eeq
$s\in S_e$, correspond to electric $p$-branes and
forms
\beq{1.10}
{\cal F}^s= Q_s \tau(\bar I_s),
\eeq
$s \in S_m$,
correspond to magnetic $p$-branes; $Q_s \neq 0$, $s \in S$.
Here  and in what follows
\beq{1.13a}
\bar I \equiv I_0 \setminus I, \qquad I_0 = \{1,\ldots,n \}.
\eeq

All manifolds $M_{i}$ are assumed to be oriented and
connected and  the volume $d_i$-forms
\beq{1.12}
\tau_i  \equiv \sqrt{|g^i(y_i)|}
\ dy_i^{1} \wedge \ldots \wedge dy_i^{d_i},
\eeq
and parameters
\beq{1.12a}
 \varepsilon(i)  \equiv {\rm sign}( \det (g^i_{m_i n_i})) = \pm 1
\eeq
are well-defined for all $i=  1,\ldots,n$.
Here $d_{i} =   {\rm dim} M_{i}$, $i =   1, \ldots, n$,
$D =   1 + \sum_{i =   1}^{n} d_{i}$. For any
 set $I =   \{ i_1, \ldots, i_k \} \in \Omega$, $i_1 < \ldots < i_k$,
we denote
\bear{1.13}
\tau(I) \equiv \tau_{i_1}  \wedge \ldots \wedge \tau_{i_k},
\\ \label{1.14}
M(I) \equiv M_{i_1}  \times  \ldots \times M_{i_k}, \\
\label{1.15}
d(I) \equiv {\rm dim }(M(I)) =  \sum_{i \in I} d_i, \\
\label{1.15a}
\varepsilon(I) \equiv \varepsilon(i_1) \ldots \varepsilon(i_k).
\ear

$M(I_s)$ is isomorphic to $p$-brane worldvolume manifold
(see (\ref{1.7})).

The parameters  $h_s$ appearing in the solution
satisfy the relations
\beq{1.16}
h_s = (B_{s s})^{-1},
\eeq
where
\beq{1.17}
B_{ss'} \equiv
d(I_s\cap I_{s'})+\frac{d(I_s)d(I_{s'})}{2-D}+
\chi_s\chi_{s'}\lambda_{\alpha a_s}\lambda_{\beta a_{s'}}
h^{\alpha\beta},
\eeq
$s, s' \in S$, with $(h^{\alpha\beta})=(h_{\alpha\beta})^{-1}$.

Here we assume that
\beq{1.17a}
({\bf i}) \qquad B_{ss} \neq 0,
\eeq
for all $s \in S$, and
\beq{1.18b}
({\bf ii}) \qquad {\rm det}(B_{s s'}) \neq 0,
\eeq
i.e. the matrix $(B_{ss'})$ is a non-degenerate one. In (\ref{1.9})
another non-degenerate matrix (a so-called "quasi-Cartan" matrix)
appears
\beq{1.18}
(A_{ss'}) = \left( 2 B_{s s'}/B_{s' s'} \right).
\eeq
Here  some ordering in the set $S$ is assumed.

In (\ref{1.3}), (\ref{1.4})
\beq{1.19}
f_s = \exp( - q^s),
\eeq
where $(q^s) = (q^s(u))$ is a solution to Toda-type equations
\beq{1.20}
\ddot{q^s} = -  \eps_s B_{s s} Q_s^2 \exp( \sum_{s' \in S} A_{s s'}
q^{s'} ),  \eeq
$s \in S$. Here
\beq{1.22}
\eps_s=(-\eps[g])^{(1-\chi_s)/2}\eps(I_s) \theta_{a_s},
\eeq
$s\in S$, $\eps[g]\equiv\sign\det(g_{MN})$. More explicitly
(\ref{1.22}) reads: $\eps_s=\eps(I_s) \theta_{a_s}$ for
$v_s = e$ and $\eps_s=-\eps[g] \eps(I_s) \theta_{a_s}$, for
$v_s = m$.

Vectors $c=(c^A)= (c^i, c^\alpha)$ and
$\bar c=(\bar c^A)$ obey the following constraints
\beq{1.27}
\sum_{i \in I_s}d_ic^i-\chi_s\lambda_{a_s\alpha}c^\alpha=0,
\qquad
\sum_{i\in I_s}d_i\bar c^i-
\chi_s\lambda_{a_s\alpha}\bar c^\alpha=0, \quad s \in S,
\eeq
\bear{1.30aa}
c^0 = \sum_{j=1}^n d_j c^j,
\qquad
\bar  c^0 = \sum_{j=1}^n d_j \bar c^j,
\\  \label{1.30a}
2E = 2E_{TL} +
 h_{\alpha\beta}c^\alpha c^\beta+ \sum_{i=1}^n d_i(c^i)^2
- \left(\sum_{i=1}^nd_ic^i\right)^2 = 0,
\ear
where
\beq{1.31}
E_{TL} = \frac{1}{4}  \sum_{s,s' \in S} h_s
A_{s s'} \dot{q^s} \dot{q^{s'}}
  + \frac{1}{2} \sum_{s \in S} \eps_s Q_s^2
  \exp( \sum_{s' \in S} A_{s s'} q^{s'} )
\eeq
is an integration constant (energy) for the solutions from
(\ref{1.20}).

Eqs. (\ref{1.20}) correspond to the
Toda-type Lagrangian
\beq{1.31a}
L_{TL} = \frac{1}{4}  \sum_{s,s' \in S}
h_s  A_{s s'} \dot{q^s}\dot{q^{s'}}
-  \frac{1}{2} \sum_{s \in S}  \eps_s Q_s^2 \exp( \sum_{s' \in S} A_{s s'}
q^{s'} ).  \eeq

The reduction of multidimensional cosmology
with composite $p$-branes (and block-diagonal metric)
to Toda-like system was performed earlier in \cite{IMJ}.

Here we identify notations  for $g^{i}$  and  $\hat{g}^{i}$, where
$\hat{g}^{i} = p_{i}^{*} g^{i}$ is the
pullback of the metric $g^{i}$  to the manifold  $M$ by the
canonical projection: $p_{i} : M \rightarrow  M_{i}$, $i = 1,
\ldots, n$. An analogous agreement will be also kept for volume forms etc.

Due to (\ref{1.9}) and  (\ref{1.10}), the dimension of
$p$-brane worldvolume $d(I_s)$ is defined by
\beq{1.16a}
d(I_s)=  n_{a_s}-1, \quad d(I_s)=   D- n_{a_s} -1,
\eeq
for $s \in S_e, S_m$, respectively.
For a $p$-brane we have $p =   p_s =   d(I_s)-1$.

\subsection{Solutions with one curved Einstein space and $n-1$
Ricci-flat spaces}

The cosmological solution with Ricci-flat spaces
may be also  modified to the following case:
\beq{1.2.2}
{\rm Ric}[g^1] = \xi_1 g^1, \  \xi_1 \ne0, \qquad {\rm Ric}[g^i] = 0, \ i
>1,
\eeq
i.e. the first space $(M_1,g^1)$ is  Einstein space of non-zero
scalar curvature and
other  spaces $(M_i,g^i)$ are Ricci-flat and
 \beq{1.2.3}
 1 \notin I_s,  \quad  \forall s  \in S,
\eeq
i.e. all ``brane'' submanifolds  do not  contain $M_1$.

In this case the exact solution may be obtained by a little modifications
of the solutions from the previous subsection \cite{IK}.

The metric reads as follows
\bear{1.2.4}
g= \biggl(\prod_{s \in S} [f_s(u)]^{2 d(I_s) h_s/(D-2)} \biggr)
\biggl\{[f_1(u-u_1)]^{2d_1/(1-d_1)}\exp(2c^1u + 2 \bar c^1) \quad
\\ \nn
\times[w du \otimes du+ f_1^2(u-u_1)g^1] +
\sum_{i = 2}^{n} \Bigl(\prod_{s\in S}
[f_s(u)]^{- 2 h_s  \delta_{i I_s} } \Bigr)\exp(2c^i u+ 2 \bar c^i)
g^i\biggr\}.
\ear
where
\bear{1.2.5}
f_1(\tau) =R \sh(\sqrt{C_1}\tau), \ C_1>0, \ \xi_1 w>0;
\\ \label{1.2.6}
R \sin(\sqrt{|C_1|}\tau), \ C_1<0, \  \xi_1 w>0;   \\ \label{1.2.7}
R \ch(\sqrt{C_1}\tau),  \ C_1>0, \ \xi_1w <0; \\ \label{1.2.8}
\left|\xi_1(d_1-1)\right|^{1/2} \tau, \ C_1=0,  \ \xi_1w>0,
\ear
$u_1$ and $C_1$ are constants, $R =  |\xi_1(d_1-1)/C_1|^{1/2}$, and
\beq{1.2.9}
C_1\frac{d_1}{d_1-1}= 2E,
\eeq
where $E$ is defined in  (\ref{1.30a}).

Now, vectors $c=(c^A)$ and $\bar c=(\bar c^A)$ satisfy
also additional constraints
\bear{1.2.10}
c^1 = \sum_{j=1}^nd_jc^j = c^0, \qquad
\bar c^1= \sum_{j=1}^nd_j\bar c^j = \bar c^0.
\ear

All other  relations from the previous subsection are unchanged.

\subsection{Restrictions on $p$-brane configurations.}

The solutions  presented above are valid if two
restrictions on the sets of $p$-branes are satisfied \cite{IK}.
These restrictions
guarantee  the block-diagonal form of the  energy-momentum tensor
and the existence of the sigma-model representation (without additional
constraints) \cite{IMC}.

Let us denote
$w_1\equiv\{i|i\in\{1,\dots,n\},\quad d_i=1\}$, and
$n_1=|w_1|$ (i.e. $n_1$ is the number of 1-dimensional spaces among
$M_i$, $i=1,\dots,n$).

{\bf Restriction 1.} {\em For any $a\in\tri$ and $v= e,m$ there are no
$I,J \in\Omega_{a,v}$ such that
$ I= \{i\} \sqcup (I \cap J)$, and $J= (I \cap J) \sqcup \{ j \}$
for some $i,j \in w_1$, $i \neq j$.}

{\bf Restriction 2.}.
{\em For any $a\in\tri$ there are no
$I\in\Omega_{a,e}$ and $J\in\Omega_{a,m}$ such that
$\bar J=\{i\}\sqcup I$.}

Restriction 1  is satisfied for $n_1 \leq 1$ and also in
the non-composite case: $|\Omega_{a,e}|+ |\Omega_{a,m}| = 1$ for all
$a\in\tri$.  For $n_1\ge2$ it forbids the following
pairs of two electric or two magnetic $p$-branes,
corresponding to the same form $F^a, a \in \tri$:

\begin{center}
\begin{tabular}{cccc}
\cline{1-2}
\multicolumn{1}{|c|}{$i$} &
\multicolumn{1}{|c|}{\hspace*{1cm}} & & $\quad I$ \\
\cline{1-2}
 & & & \\
\cline{2-3}
 & \multicolumn{1}{|c|}{\hspace*{1cm}} &
\multicolumn{1}{|c|}{$j$} & $\quad J$ \\
\cline{2-3}
\end{tabular}
\end{center}

\begin{center}

{\bf \small Figure 1.
A forbidden by Restriction 1 pair of two electric or two
magnetic p-branes.
}

\end{center}

Here $d_i = d_j =1$, $i \neq j$, $i,j =1,\dots,n$. Restriction 1
may be also rewritten in terms of intersections
\beq{1.3.1a}
{\bf (R1)} \quad d(I \cap J) \leq d(I)  - 2,
\eeq
for any $I,J \in\Omega_{a,v}$, $a\in\tri$, $v= e,m$ (here $d(I) = d(J)$).

Restriction 2 is satisfied for $n_1=0$. For
$n_1\ge1$ it forbids the following electro-magnetic pairs,
corresponding to the same form $F^a, a \in \tri$:

\begin{center}
\begin{tabular}{ccc}
\cline{1-2}
\multicolumn{1}{|c|}{\hspace*{1cm}} &
\multicolumn{1}{|c|}{$i$}  & $\quad\bar J$ \\
\cline{1-2}
 & & \\
\cline{1-1}
\multicolumn{1}{|c|}{\hspace*{1cm}} & & $\quad I$
\\
\cline{1-1}
\end{tabular}
\end{center}

\begin{center}

{\bf \small Figure 2.
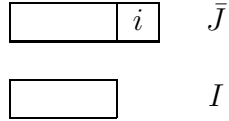
Forbidden by Restriction 2 electromagnetic pair of
p-branes}

\end{center}

Here $d_i =1$, $i =1,\dots,n$. In terms of intersections
Restriction 2 reads
\beq{1.3.1b}
{\bf (R2)} \quad d(I \cap J) \neq 0,
\eeq
for $I\in\Omega_{a,e}$ and $J\in\Omega_{a,m}$, $a \in \tri$.

{\bf Intersection rules.}
>From  (\ref{1.16}), (\ref{1.17}) and (\ref{1.18})  we get
the  $p$-brane intersection rules  corresponding
to the quasi-Cartan matrix $(A_{s s'})$ \cite{IMJ}
\beq{1.3.4}
d(I_s \cap I_{s'})= \frac{d(I_s)d(I_{s'})}{D-2}-
\chi_s\chi_{s'}\lambda_{a_s}\cdot\lambda_{a_{s'}} + \frac12 B_{s's'}
A_{ss'}, \eeq
where $\lambda_{a_s}\cdot\lambda_{a_{s'}} =
\lambda_{\alpha a_s}\lambda_{\beta a_{s'}} h^{\alpha\beta}$;
$s, s' \in S$.

\section{Composite S-brane solutions}

In this section we single out special
cosmological solutions called as $S$-branes.

\subsection{Toda-type solutions}

In what follows we suppose
that all  metrics $g^i$ have Euclidean signatures $(+,...,+)$
and  $w = -1$.
Thus the total metric $g$ has the pseudo-Euclidean signature  $(-,+,...,+)$.
We put in  action (\ref{1.1})  $\theta_a = 1$.

Then, from definitions (\ref{1.15a}) and (\ref{1.22}) we get
\beq{2.2}
\eps_s=\eps(I_s) =1
\eeq
for all $s$.

We also put
\beq{2.3}
d_1 > 1
\eeq
and $1 \notin I_s$, i.e. all branes do not contain $M_1$-submanifold.

Let us assume that
the (brane) matrix $(B_{ss'})$ is a positive definite
and hence
\beq{2.5}
 B_{ss} > 0
\eeq
for all $s \in S$.
This suggestion and  (\ref{2.2}) imply
\beq{2.6}
E_{TL} > 0.
\eeq

Now we make  the following choice of parameters in the solutions
>from subsections 2.2 and 2.3
\beq{2.7}
c^i = 0, \ i >1, \qquad c^\alpha=0,
\eeq
for all $\alpha$ and
\beq {2.7a}
\bar c^i = \bar c^\alpha = 0,
\eeq
for all $i, \alpha$.
With this choice the brane constraints (\ref{1.27}) are satisfied
identically due to  (\ref{1.2.3}).

With the adopted choice of parameters both
 solutions from
subsections 2.2 and 2.3 may be unified by the following relations
\bear{2.8a}
g= \biggl(\prod_{s \in S} H_s^{2 d(I_s) h_s/(D-2)} \biggr)
\biggl\{H^{2d_1/(1-d_1)} [ - dt \otimes dt+ H^2 g^1] \\ \nn
+ \sum_{i = 2}^{n} \Bigl(\prod_{s\in S}
H_s^{- 2 h_s  \delta_{i I_s} } \Bigr) g^i\biggr\},
\\ \label{2.8b}
\exp(\varphi^\alpha) =
\prod_{s\in S} H_s^{h_s \chi_s \lambda_{a_s}^\alpha},
\\ \label{2.8c}
F^a= \sum_{s \in S_e} \delta^a_{a_s} Q_s
\Bigl( \prod_{s' \in S}  H_{s'}^{- A_{s s'}} \Bigr) dt \wedge\tau(I_s)
+ \sum_{s \in S_m}\delta^a_{a_s} Q_s \tau(\bar I_s),
\ear
$a \in \tri$; $\alpha = 1, \ldots, l$;
where
\bear{2.9a}
H = |\xi_1(d_1-1)|^{1/2}  \frac{\sh(M t)}{M},  \ \xi_1 > 0;
\\ \label{2.9b}
|\xi_1(d_1-1)|^{1/2} \frac{\ch(M t)}{M},  \ \xi_1 < 0;  \\
 \label{2.9c}
\exp( M t),  \ \xi_1  = 0;
\ear
$M > 0$, and
\beq{2.10}
\frac{M^2 d_1}{d_1-1}= 2E_{TL} = \frac{1}{2}  \sum_{s,s' \in S} h_s
A_{s s'} \dot{q^s} \dot{q^{s'}}
  +  \sum_{s \in S}  Q_s^2 \exp( \sum_{s' \in S} A_{s s'} q^{s'} ).
\eeq

Here
\beq{2.11}
H_s  = \exp( - q^s(t)),
\eeq
with $q^s(t)$ obeying Toda-type equations
\beq{2.12}
\ddot{q^s} = -   B_{s s} Q_s^2 \exp( \sum_{s' \in S} A_{s s'}
q^{s'} ), \quad s \in S.
\eeq

The metric  $g^1$ is an Einstein metric on $M_1$:
${\rm Ric}[g^1] = \xi_1 g^1$, and all other metrics are Ricci-flat, i.e.
${\rm Ric }[g^i] = 0$ for $i > 1$.

In combining of two subfamilies of solutions from the previous section
we used the following definitions and identifications: $u = t$,
$f_1 = H$, $f_s = H_s$, $M = \sqrt{C_1}$ ($C_1 > 0$) for $\xi_1 \neq 0$ and
$(d_1 -1) c^1 = M$ for $\xi_1 = 0$.

\subsection{Solutions with "orthogonal" intersections}

Let us consider "orthogonal" (or $A_1 \oplus ...\oplus A_1$) intersection
 rules
 \beq{2.13} B_{s s'}=0, \quad s \neq s', \eeq
 (remind that
 $B_{s s'} = (U^{s},U^{s'})$
 are scalar products of brane vectors \cite{IMC,IMtop}).

Then using relations from Appendix we get
 \beq{2.14}
H_s = |Q_s||h_{s}|^{-1/2}  \frac{\ch(M_s (t-t_s))}{M_s},
\eeq
where all $M_s > 0$  and $t_s$ are constants and
\beq{2.15}
2 E_{TL} =  \sum_{s \in S} M_s^2 h_s.
\eeq

Here we used the notations $M_s =\sqrt{C_s}$, $C_s > 0$ (see Appendix).
For $\xi_1 \neq 0$ these "orthogonal" $S$-brane solutions are special cases
of more general ones from \cite{IMJ}.

\subsection{Solutions with "block-orthogonal" intersections}

Let us suppose that
\beq{2.13a}
B_{s s'}=0, \quad  s \in S_i, \ s'\in S_j, \ i \neq j,
\eeq
where
\beq{2.13b}
S=S_1\sqcup\dots\sqcup S_k,
\eeq
$S_i\ne\emptyset$, $i,j=1,\dots,k$.
Relation (\ref{2.13b}) means that the set $S$ splits into  $k$
mutually non-intersecting subsets (blocks) $S_1,\dots,S_k$.

Then using relations from Appendix we get
 \beq{2.14b}
H_s = \left[ |Q_s||h_s b_s|^{-1/2}  \frac{\ch(M_s (t-t_s))}{M_s}
\right]^{b_s},
 \eeq
where all $M_s > 0$ and $t_s$ are constants
coinciding inside "blocks":
\beq{2.14c}
t_s = t_{s'}, \quad M_s = M_{s'},
\eeq
$s,s' \in S_i$, $i = 1, \ldots, k$.
The charges should be proportional to each other inside blocks
\bear{1.4.10b}
\frac{Q_s^2}{b_s h_s} = \frac{Q_{s'}^2}{b_{s'} h_{s'}},
\ear
$s,s' \in S_i$, $i = 1, \ldots, k$.
Here
\beq{1.4.4}
b_s = 2 \sum_{s' \in S} A^{s s'}, \quad (A^{ss'})= (A_{ss'})^{-1}.
\eeq

For the Toda-like part of energy we get from Appendix
\beq{2.15b}
2 E_{TL} =  \sum_{s \in S} M_s^2 b_s h_s.
\eeq

More general "block-orthogonal" solutions where considered in
\cite{Br1,IMJ1,IMJ2}.

%%%%%%%%%%%%%%%%%%%%%%%%%%%%%%%%%%%%%%%%%%%%%%%%%%%%%%%%%%%%%%%%
\subsection{Solutions related to $A_m$  Toda  chain}
%%%%%%%%%%%%%%%%%%%%%%%%%%%%%%%%%%%%%%%%%%%%%%%%%%%%%%%%%%%%%%%%

Here we  consider exact solutions to
Toda-chain  (\ref{2.12}) equations \cite{T}
corresponding to the Lie algebra
 $A_m= sl(m+1, \C)$
with the Cartan matrix
\beq{B.1a}
\left(A_{ss'}\right)=
\left( \begin{array}{*{6}{c}}
2&-1&0&\ldots&0&0\\
-1&2&-1&\ldots&0&0\\
0&-1&2&\ldots&0&0\\
\multicolumn{6}{c}{\dotfill}\\
0&0&0&\ldots&2&-1\\
0&0&0&\ldots&-1&2
\end{array}
\right)\quad
\eeq
$s,s' \in S =  \{1, \ldots, m \}$.

The Toda chain equations have the following solutions \cite{And,GM,IK}
\beq{B.3}
H_s =
C_s^{-1} \sum_{r_1< \dots <r_s}^{m+1} v_{r_1}\cdots v_{r_s}
\Delta^2( w_{r_1}, \ldots, w_{r_s}) \exp[(w_{r_1}+\ldots +w_{r_s}) t],
\eeq
$s = 1, \ldots, m$, where
\beq{B.4a}
\Delta( w_{r_1}, \ldots, w_{r_s})  =
\prod_{i<j}^{s} \left(w_{r_i}-w_{r_j}\right); \quad
\Delta(w_{r_1}) \equiv 1,
\eeq
denotes the  Vandermonde determinant.
The real constants $v_r$ and $w_r$, $r = 1, \ldots, m + 1$, obey the
relations
\beq{B.5}
\prod_{r=1}^{m+1} v_r= \Delta^{-2}(w_1,\ldots, w_{m+1}), \qquad
\sum_{r=1}^{m+1}w_r=0.
\eeq
In (\ref{B.3})
\beq{B.6}
C_s = \prod_{s'=1}^{m} [B_{s's'} Q_{s'}^2]^{-A^{s s'}},
\eeq
where
\beq{B.7}
A^{s s'}= \frac1{m+1}\min(s,s')[m+1-\max(s,s')],
\eeq
$s, s' = 1, \ldots, m$,  \cite{FS}.
Here $v_r \neq 0$ and $w_r \neq w_{r'}$, $r \neq r'$;
$r, r' = 1, \ldots, m +1$.
Due to  $B_{ss} > 0$, $s \in S$,  all $w_r, v_r$ are real, and, moreover,
all  $v_r > 0$, $r =1, \ldots, m+1$ \cite{IK}.

For the (Toda) energy  we get
\beq{B.10}
E_{TL} =  \frac{h}{4} \sum_{r=1}^{m+1} w^2_r,
\eeq
where $h = (B_{ss})^{-1}$,
$s \in S$ (here all $B_{ss}$ are equal).

It should be noted that pioneering
cosmological solutions with $p$-branes related to
Toda chains were considered earlier in  \cite{LPX,LMPX,LP1}.

\section{Examples}

Here we present well-known $SM$-brane solutions in $D =11$ supergravity
\cite{CJS}.

\subsection{Solutions for algebra $A_1$}

We start with single $S$-branes.

\subsubsection{$SM2$-brane}

Let $n =3$, $d_2 = 3$.
The $SM2$-brane solution reads
\bear{4.1}
g=  H^{1/3}_1  \biggl\{H^{2d_1/(1-d_1)}[ - dt \otimes dt+ H^2 g^1]
+  H_1^{-1} g^2 + g^3   \biggr\},
\\  \label{4.1a}
F= Q_1 H^{- 2}_1 dt  \wedge \tau_2,
\ear
where
$H$ is defined in (\ref{2.9a})-(\ref{2.9c}),
$H_1$ is defined in  (\ref{2.14}) ($h_{1} =1/2$)
and $2M^2 d_1/(d_1-1)=  M_1^2$.

\subsubsection{$SM5$-brane}

Let us consider the magnetic solution dual to $SM2$.
We put $n =3$, $d_2 = 6$.
The solution reads
\bear{4.2b}
g=  H^{2/3}_1  \biggl\{H^{2d_1/(1-d_1)}[ - dt \otimes dt+ H^2
g^1]
+  H_1^{-1} g^2 + g^3   \biggr\},
\\  \label{4.2ab}
F= Q_1 \tau_1 \wedge \tau_3 ,
\ear
where $H$ and $H_1$ are defined in (\ref{2.9a})-(\ref{2.9c})
and (\ref{2.14}),
respectively,  ($h_{1} =1/2$) and $2M^2 d_1/(d_1-1)=  M_1^2$.

\subsection{$SM2 \cap SM5$-branes with $A_1 \oplus A_1$ intersection}

Here we present a "superposition" of $SM2$ and $SM5$
solutions corresponding to "ortogonal intersection":
$d(I_1 \cap I_2) =2$.
We put $n =5$, $d_1 =2$, $d_2 = 1$, $d_3 = 2$, $d_4 = 4$, $d_5 = 1$
and $I_1 = \{ 2,3 \}$, $I_2 = \{ 3, 4 \}$.
The solution reads
\bear{4.3}
g= H_1^{1/3}  H_2^{2/3}  \biggl\{H^{-4}[ - dt \otimes dt+ H^2 g^1]
+  H_1^{-1} g^2 +
\\ \nn H_1^{-1} H_2^{-1} g^3 + H_2^{-1} g^4 +   g^5   \biggr\},
\\  \label{4.3a}
F= Q_1 H_1^{- 2}dt  \wedge \tau_2 \wedge \tau_3 + Q_2 \tau_1 \wedge \tau_2
\wedge \tau_5,  \ear
where $H$ and $H_s$ are  defined in  (\ref{2.9a})-(\ref{2.9c}) and
(\ref{2.14}), respectively (all $h_{s} =1/2$) and $4M^2 =  M_1^2 + M_2^2$.

\subsection{$SM2 \cap SM5$-branes with $A_2$ intersection}

Now we consider a (dyonic) solution consisting of  $SM2$ and
$SM5$ branes and with $A_2$ intersection: $d(I_1 \cap I_2) =1$ \cite{IMtop}.
We put $n =4$, $d_1 = d_2 = 2$, $d_3 = 1$, $d_4 = 5$
and $I_1 = \{ 2,3 \}$, $I_2 = \{ 3, 4 \}$.
The  solution reads (see subsection 3.3)
\bear{4.4}
g= H_1  \biggl\{H^{-4}[ - dt \otimes dt+ H^2 g^1]
+  H_1^{-1} g^2 +
\\ \nn H_1^{-2}  g^3 + H_1^{-1} g^4  \biggr\},
\\  \label{4.4a}
F= Q_1 H_1^{- 1}dt  \wedge \tau_2 \wedge \tau_3 \pm Q_1 \tau_1 \wedge
\tau_2,  \ear
where $H$ is defined in  (\ref{2.9a})-(\ref{2.9c}) and
 \beq{4.4b}
H_1 = \left[ |Q_1|  \frac{\ch(M (t-t_1))}{M}\right]^{2}.
 \eeq

This solution is also a special case of $A_2$ Toda chain
solution (with $Q_1^2 = Q_2^2$) from subsection 3.4.

\section{Kasner-like asymptotical behaviour}

Here we consider asymptotical behaviour of  $S$-brane solutions
in the limit $t \to +\infty$, when
\beq{s.1}
H \sim {\rm const} \exp(Mt), \qquad  H_s \sim {\rm const} \exp(M_s t),
\eeq
$s \in S$.

It may be verified by a straightforward calculation for one-brane
case that  in the limit $t \to +\infty$
the metric and scalar fields have asymptotical
Kasner-like behaviour
\bear{s.2}
g_{as}=
- d\tau \otimes d\tau + \sum_{i =1}^{n} A_i \tau^{2 \alpha^i}g^i, \\
\varphi^\beta_{as}  = \alpha^{\beta} \ln \tau + const,
\ear
as $\tau \to +0$ ($\tau$ is synchronous time variable)
with  the set of Kasner parameters $\alpha =
(\alpha^{i}, \alpha^{\gamma})$, obeying
\beq{s.3}
\sum_{i=1}^{n} d_i \alpha^i =
\sum_{i=1}^{n} d_i (\alpha^i)^2 +
\alpha^{\beta} \alpha^{\gamma} h_{\beta \gamma}= 1.
\eeq
Here $\tau \sim {\rm const} \exp(- {\cal M}t)$ for $t \to +\infty$,
where ${\cal M} > 0$ is a linear combination of "masses" $M$, $M_s$.
Analogous (though more tedious) calculations for several branes with
orthogonal intersections give an analogous result.

Here we show that the  Kasner-like asymptotical behaviour may be proved
using the billiard representation for multidimensional cosmology with
branes \cite{IMb1} under  restrictions imposed
in Section 3 and the following additional
condition added: the  matrix $(h_{\alpha\beta})$ is supposed to be
positive definite.
According to results of refs. \cite{IMb1}
for the proof of the Kasner-like asymptotical
behaviour it is sufficient to verify
that there exists a Kasner set obeying the  inequalities
\beq{5.1}
U^s(\alpha) =   \sum_{i\in I_s}d_i\alpha^i
 -\chi_s\lambda_{a_s \gamma}\alpha^{\gamma} > 0,
\eeq
$s \in S$.
Indeed, let us consider the following set  of Kasner parameters
\bear{5.2}
 \alpha^1 =   (d_1 - \Delta)/[d_1(D-1)], \\
\alpha^2 =...= \alpha^n =  (d + \Delta)/[d(D-1)], \\
\alpha^{\gamma} = 0,
\ear
where $d = d_2 +...+ d_n$ and $\Delta = \sqrt{d_1 d (D-2)}$. Due to
condition $1 \notin I_s$ this Kasner set  does satisfy the relations
(\ref{5.1}).

When all $d_i > 1$ and
$\alpha^{\beta} \alpha^{\gamma} h_{\beta \gamma} = 0$,
then the Riemann tensor squared of the Kasner-type metric
(\ref{s.2}) diverges for $\tau \to +0$ \cite{IMsing} and, hence,
the Riemann tensor squared of the original metric $g$
diverges as $t \to + \infty$.
For  $\alpha^{\beta} \alpha^{\gamma} h_{\beta \gamma} \neq 0$
the scalar curvature of (\ref{s.2}) diverges as $\tau \to +0$
that implies $R[g] \to \infty$  for $t \to + \infty$.

Analogous consideration may be carried out for the solution
with $H$ from (\ref{2.9b}) in the limit  $t \to -\infty$.

{\bf Remark. Conical singularity.}
Let us consider  the asymptotical behaviour of the
solution with $H$ from
(\ref{2.9a}) when  $t \to +0$. In general case we get
a conical singularity in this limit. This singularity
is absent in special case when
$(M_1,g^1)$ is unit Lobachevsky space of dimension
$d_1$ (and $\xi_1 = -(d_1 -1)$).
This is a well-known Milne-type resolution of singularity.

{\bf Remark. $S$-brane cosmology with oscillating
behaviour near the singularity.}
In  the pioneering paper on billiard approach
for multidimensional cosmology
with $p$-branes \cite{IMb1} a large variety
of composite $S$-brane solutions with never-ending
oscillating behaviour near the singularity
was described (and certain examples were considered).
The open problem is to find examples of such type solutions
(with block-orthogonal metrics) in string cosmology.
We note that recent results  of Damour and Henneaux
on chaotic behaviour in  $D=10,11$ supergravities
\cite{DamH1,DamH2} (based on inequalities (\ref{5.1}))
deal with  $1$-dimensional manifolds $M_1 = \dots = M_n$
and sets of (composite) electric branes with "forbidden" intersections
(see subsection 2.4) that lead to additional constraints
\cite{IMC}. The recent mathematical analysis carried out
in the paper \cite{DHRW} gives
certain arguments in favour of asymptotically
block-orthogonal (but not obviously block-orthogonal) form of metric
in the vicinity of singularity
for pure cosmological solutions in string cosmology with
composite $p$-brane configurations from \cite{DamH1,DamH2}.

%%%%%%%%%%%%%%%%%%%%%%%%%%%%%%%%%%%%%%%%%%%%%%%%%%%%%%%%%%%
\section{Conclusions and discussions}
%%%%%%%%%%%%%%%%%%%%%%%%%%%%%%%%%%%%%%%%%%%%%%%%%%%%%%%%%%%

In this paper  a family of Toda-like
composite $S$-brane solutions was presented.
These ($S$-brane) solutions are special
case of more general cosmological-type solutions from  \cite{IK}.
They generalize $S$-brane solutions from refs. \cite{S1,S2,S3,S4,S5}
to a composite  configurations with next to arbitrary
intersecting rules.

Here  several subclasses of $S$-brane solutions
with "orthogonal",  "block-orthogonal"
and $A_m$ intersections were considered and  certain examples  of
solutions in $D= 11$ supergravity (describing $SM$-branes with
$A_1$, $A_1 \oplus A_1$ and  $A_2$ intersections),  were singled out.
%(For non-Ricci-flat space $(M_1,g^1)$ general
%classical and quantum $S$-brane solutions with
%"orthogonal" intersection rules were considered earlier in \cite{IMJ}.)

The solutions under consideration have (in general position) an
asymptotical  Kasner-like behaviour near the singularity when $t \to
+ \infty$.  This fact was shown here using the billiard approach
for $p$-brane cosmology \cite{IMb1} under the assumption
$1 \notin I_s$ imposed (i.e.
when all branes do not contain $M_1$-submanifold).
The relaxing of this assumption  may lead
to $S$-brane configurations with oscillating
behaviour near the singularity (see  \cite{IMb1}).

We remind that in refs. \cite{IMJ,IMAJ}  the Wheeler-DeWitt (WDW) equation
for the quantum cosmology with composite electro-magnetic $p$-branes
(e.g. $S$-branes) defined on product of Einstein spaces
was obtained (for non-composite  electric case see also \cite{GrIM}).
In these papers the WDW equation
was integrated for intersecting $p$-branes with orthogonal $U$-vectors,
when $n-1$ internal spaces are Ricci-flat and one is
the Einstein space of a
non-zero curvature (for non-composite electric case see \cite{GrIM}).
A slightly different approach with classical field of forms (and
a special brane setup) was suggested in \cite{LMMP}.
An open problem is to generalize classical
and quantum $S$-brane solutions  from \cite{IMJ}
(corresponding to the Lie algebra $A_1 \oplus \dots \oplus A_1$)
to other semisimple Lie algebras.

%%%%%%%%%%%%%%%%%%%%%%%%%%%%%%%%%%%%%%%%%%%%%%%%%%%%%%%%%%%%%%%%%%%%%%%%

\newpage

\begin{center}
{\bf Acknowledgments}
\end{center}

This work was supported in part by the Russian Ministry of
Science and Technology, Russian Foundation for Basic Research
(RFFI-01-02-17312-a) and Project DFG (436 RUS 113/678/0-1(R)).
The author thanks Prof. H. Dehnen
and colleagues from Physical Department of the University of Konstanz for
hospitality during his visit in May-July 2002.

\vspace{10pt}

{\large \bf Appendix}

\renewcommand{\theequation}{A.\arabic{equation}}

\vspace{10pt}

{\bf Solutions for "block-orthogonal" intersections.}
Here we consider "block-orthogonal" case (\ref{2.13a}), (\ref{2.13b}).
In this  case the moduli functions (\ref{1.19})
read  \cite{IMJ2,IMJ1,IMtop}
\beq{1.4.3}
f_s = (\bar{f}_s)^{b_s},
\eeq
with powers $b_s$ defined in (\ref{1.4.4}) and
\bear{1.4.5}
\bar{f}_s(u)=
R_s \sh(\sqrt{C_s}(u-u_s)), \;
C_s>0, \; \eta_s\eps_s<0; \\ \label{1.4.7}
R_s \sin(\sqrt{|C_s|}(u-u_s)), \;
C_s<0, \; \eta_s\eps_s<0; \\ \label{1.4.8}
R_s \ch(\sqrt{C_s}(u-u_s)), \;
C_s>0, \; \eta_s\eps_s>0; \\ \label{1.4.9}
|Q^s||b_s h_s|^{-1/2}(u-u_s), \; C_s=0, \; \eta_s\eps_s<0,
\ear
where $R_s = |Q_s||b_s h_s C_s|^{-1/2}$,
$\eta_s = {\rm sign}(b_s h_s) = \pm 1$,
$C_s$, $u_s$ -- are constants, $s \in S$, coinciding
inside "blocks":
$u_s = u_{s'}$, $C_s = C_{s'}$,
$s,s' \in S_i$; $i = 1, \ldots, k$.

The charges should be proportional to each other inside blocks
\bear{1.4.10}
\frac{\eps_s Q_s^2}{b_s h_s} = \frac{\eps_{s'} Q_{s'}^2}{b_{s'} h_{s'}},
\ear
$s,s' \in S_i$, $i = 1, \ldots, k$.

The Toda part of the energy reads in this case
\beq{1.4.11}
E_{TL} = \frac{1}{2} \sum_{s \in S} C_s b_s h_s.
\eeq

{\bf "Orthogonal" case. }
In the "orthogonal" case when all blocks consist of one element,
i.e.  $|S_1| = \ldots = |S_k| = 1$ \cite{IMJ},
the quasy-Cartan matrix is diagonal   $A = {\rm
 diag}(2, \ldots,2)$, all $b_s = 1$
and
\beq{1.4.3a}
f_s = \bar{f}_s,
\eeq
$s \in S$.

\end{document}